\documentclass[12pt]{article}
\usepackage{geometry}  \usepackage
{ifthen,color,xypic,amssymb}
\usepackage{graphicx}
\usepackage{amsmath}
\usepackage{latexsym}

\def\x{\times}
\def\ox{\otimes}
\def\o+{\oplus}
\def\ra{\rightarrow}

\def\lra{\longrightarrow}

\def\beqa{\begin{eqnarray}}
\def\eeqa{\end{eqnarray}}

\newcommand{\al}{\alpha}

\newcommand{\la}{\lambda}
\newcommand{\si}{\sigma}

\newcommand{\om}{\omega}
\newcommand{\C}{{\cal C}}

\newcommand{\cO}{{\cal O}}

\newcommand{\resetcounter}{\setcounter{equation}{0}}

\begin{document}
\thispagestyle{empty}
\rightline{LMU-ASC 77/06}
\rightline{hep-th/0611309}
\vspace{2truecm}
\centerline{\LARGE Heterotic Models without Fivebranes}
\vspace{0.2truecm}

\vspace{1.5truecm}
\centerline{Bj\"orn Andreas$^1$ and Gottfried Curio$^2$}

\vspace{.6truecm}
\centerline{$^1${\em Institut f\"ur Mathematik und Informatik,
Freie Universit\"at Berlin}}
\centerline{\em Arnimallee 14, 14195 Berlin, Germany}
\centerline{$^2${\em Arnold-Sommerfeld-Center for Theoretical Physics}}
\centerline{{\em Department f\"ur Physik,
Ludwig-Maximilians-Universit\"at M\"unchen}}
\centerline{{\em Theresienstr. 37, 80333 M\"unchen, Germany}}

\vspace{1.0truecm}

\begin{abstract}

After discussing some general problems for 
heterotic compactifications involving fivebranes
we construct bundles, built as extensions, 
over an elliptically fibered Calabi-Yau threefold. 
For these we show that it is possible to satisfy 
the anomaly cancellation topologically without any fivebranes. 
The search for a specific Standard model or GUT gauge group
motivates the choice of an Enriques surface 
or certain other surfaces as base manifold.
The burden of this construction is to show the stability of these bundles.
Here we give an outline of the construction and its physical relevance.
The mathematical details, in particular the proof that the bundles are stable
in a specific region of the K\"ahler cone,
are given in the mathematical companion paper math.AG/0611762. 

\end{abstract}

\newpage

\section{On problems caused by fivebranes and the $H$-field}

Let us recall first some problems with fivebranes and the $H$-field
which, although known in principle, were sometimes neglected
in the literature.
These problems lead to some difficulties in the 
usage of bundles $V$ (say in the observable sector) over a
Calabi-Yau space $X$ for a heterotic string compactification. 
Apart from the standard embedding $F=R$
the anomaly cancellation condition leads in general (already when read 
just on a cohomological level) to the occurrence of a fivebrane class $W$.
We argue that this enforces the existence of a non-vanishing $H$-field
which is of a markedly singular character. 
Even when it is possible to avoid this
(for example by interpreting this class $W$
as the $c_2(V_{hid})$ of a hidden bundle) 
one encounters still the same problematic occurrence of the $H$-field:
this is when one realizes that the anomaly condition 
has actually to be solved on the form level already.
A non-trivial $H$-field however is known to lead, via supersymmetry, 
to compatibility conditions on the underlying space geometry $X$,
which turns out to be non-K\"ahler.
This has the consequence that the usual Donaldson-Uhlenbeck-Yau (DUY) 
theorem is no longer applicable. This theorem assured
the solvability of the equation of motion
$F^{a\bar{b}}J_{a\bar{b}}=\frac{1}{2}F \wedge J^2=0$ 
on the form level 
from the condition $c_1(V)J^2=0$ on the cohomological level
for bundles $V$ stable with respect to a suitable K\"ahler class $J$. 
A generalisation for the non-K\"ahler case [\ref{LY}]
gives the solvability of the equation of motion for $X$ having a
Gauduchon metric, using an appropriate stability notion.
Still one has to solve the non-linear anomaly equation exactly in a manner 
which goes beyond the 
perturbative arguments arguing by corrections order by order
[\ref{Wit}]. Recently first steps in that direction were made [\ref{Yau}].

\subsection{Problems when fivebranes are present}

Besides solving the equation of motion $F\wedge J^2=0$, 
the field strength $F$ must 
also obey the anomaly equation
\beqa
dH = \frac{\al'}{4}\big( tr\; R\wedge R - tr\; F\wedge F \big)
\eeqa
Many known vector bundle constructions, such as the spectral cover method, 
explicitly violate the ensuing topological condition $c_2(X)=c_2(V)$. 
So one needs to introduce 
a number of space-time filling fivebranes and has to use the 
generalised anomaly equation
\beqa
\label{form level}
dH = \frac{\al'}{4}\Big( tr\; R\wedge R - tr\; F\wedge F
+ 16\pi^2 \delta_W\Big)
\eeqa
Here a further four-form 
arises which is supported on the codimension four world-volume of $W$;
this is a four-form with distribution coefficients, i.e.~a current. 
The possibility of having a $W\neq 0$ 
has been first treated on a computational level in [\ref{FMW}]. 

The necessary condition for solving (\ref{form level}) is the 
corresponding integrability condition
\beqa
c_2(X)=c_2(V)+W
\eeqa
(assuming $c_1(V)=0$)
where $W$ has to be realisable by an effective class of holomorphic curves.
However, the condition actually to be solved is (\ref{form level}) 
on the form level.

The presence of fivebranes prevents heterotic models to be 
interpreted as perturbative nonlinear sigma models.
Studies of stability of $(0,2)$ models usually assume 
the absence of fivebranes. 
Especially problematic is that
the delta-function $\delta_W$ is not cancelled by 
another term on the right hand side which indicates that
one cannot have a consistent solution without an $H$-field
\beqa
W=c_2(X)-c_2(V)\neq 0 & \Longrightarrow & H\neq 0
\eeqa
The non-trivial $H$-field must fulfil even $dH\neq 0$.
More precisely, $H$ must be chosen 
such that $dH$ contains the relevant current.
The singular character of $H$ becomes especially severe
when one realizes that in (\ref{form level}) a consistent $H$-solution 
occurs simultaneously 
on the right hand side, inside the connection with torsion
$\om +aH$ from which the $tr \; R\wedge R$ - term is computed (there is 
an important issue in this framework as to which $a$ has to be
used in different places; 
for our purposes we do not need to enter this discussion).

One solution of this singularity problem, 
caused by the current contribution, could lie in dissolving
the delta-function-like fundamental-brane (given by the small instanton)
into a smooth solitonic (gauge-)field configuration.
However, although such fivebrane contributions $W$ 
can occur as singular limits 
of gauge bundles, where (a part of)
the curvature term $tr\; F\wedge F$ degenerates 
to a delta function source, we are not concerned here with 
such small instanton transitions.
These change the rank$(V)$ and/or $c_3(V)$ 
[\ref{FMW}],[\ref{db}],[\ref{ov}];
so if one insists on certain 
values for these invariants of $V_{(obs)}$ 
(say from phenomenological investigations in the obeservable sector) 
giving a $W\neq 0$,
it is of no help to change to 
another bundle $\tilde{V}_{(obs)}$ where these phenomenologically decisive 
invariants differ and just $W$ is absorbed into $c_2(\tilde{V}_{(obs)})$.\\

\noindent
One route, employed in the present paper, to avoid the problem at least
on the cohomological level is to construct a $V$ with $c_2(X)-c_2(V)=0$.
Alternatively 
the problem could be circumvented if one could solve the anomaly constraint 
with the help of a second bundle in the hidden sector 
which realises the class $W$ as $c_2(V_{hid})$ 
(cf. also [\ref{Thomas}]), leaving $V=V_{obs}$ intact. 
Thereby one stays within the framework of 
perturbative $(0,2)$ models characterised by stable holomorphic 
bundles $(V, V_{hid})$ of $c_1(V)=c_1(V_{hid})=0$ 
(this condition can be relaxed) with
\beqa
c_2(V)+c_2(V_{hid})=c_2(X)
\eeqa
(cf. in this connection also [\ref{DZ}]).
This leads therefore to the following general problem: 
\begin{itemize}
\item 
Suppose an effective holomorphic curve class
(a sum of irreducible holomorphic curves with non-negative integral
coefficients) is given\footnote{Stricly speaking 
the assumption on $W$ is more specific as it is 
$c_2(X)-c_2(V)$ for a stable bundle $V$.} 
which represents the compact support of the
(space-time filling) fivebrane and whose cohomology class is denoted by
\beqa
W\in H^4(X, {\bf Z})
\eeqa
When can $W$ be represented 
as second Chern class of a vector bundle (of $c_1(V_{hid})=0$)
\beqa
\label{problem}
W=\; c_2(V_{hid})\;\;\;\; ?
\eeqa
Here $V_{hid}$ and $V$ have to be stable with respect to the same K\"ahler 
class.

\end{itemize}
Although a precise, or at least sufficient condition is not known,
one knows a necessary condition: a holomorphic vector bundle $V_{hid}$
(stable w.r.t.~$J$) satisfies the Bogomolov
inequality
\beqa
\label{Bogo}
0 & \leq & c_2(V_{hid})J
\eeqa
The ensuing necessary condition $0\leq WJ$ is satisfied
in our case as $W$ was an effective class.

In [\ref{Douglas}] the conjecture is put forward that
a stable bundle (more precisely reflexive sheaf) $V_{hid}$ 
of rank $r_{hid}$ and\footnote{In addition $c_3(V_{hid})$ 
can be conjecturally chosen freely if it is
$ < \frac{16\sqrt{2}}{3}r_{hid}D^3$.} 
$c_1(V_{hid})=0$ 
exists if one has for some ample class $D$ that
\beqa
\label{conjecture}
c_2(V_{hid})-\frac{r_{hid}}{24}c_2(X)& = & r_{hid}D^2
\eeqa
Assuming this conjecture 
to be true (\ref{problem}) can be solved if (\ref{conjecture}) holds
for $W$. 
Proving (\ref{problem}) this way, one needs not only
$W$ effective with $0\leq WJ\leq c_2(X)J$ (the latter from
$0\leq c_2(V)J$), but even the following condition 
(note $0\leq c_2(X)J$ by the Miyaoka-Yau theorem)
\beqa
\frac{r_{hid}}{24}c_2(X)J\; \stackrel{!}{\leq} \; W\, J \; \leq \; c_2(X)J
\eeqa

\subsection{Problems when no fivebranes are present}

Avoiding fivebrane contributions is of course not the end 
of the problems for consistent heterotic string compactifications
caused by the anomaly cancellation requirement.
For this let us assume that no need arises for a current respresenting 
a fivebrane contribution (something which is 
already cohomologically detectable); 
or assume that one has succeeded to represent such a contribution 
by a bundle in the hidden sector. Then still one has the problem that 
(\ref{form level}) actually has to be solved locally, i.e., 
{\em on the form level}. If one is not in the quite exceptional
case of having an $F\neq R$ with $tr\; F\wedge F = tr\; R\wedge R$ locally,
this will necessitate to turn on of a non-trivial $H$-field
(though this time smooth and not being singular to balance by its $dH$
a delta-function contribution $\delta_W$).

However, having now a non-trivial $H$-field turned on, 
the compatibility conditions (stemming from the requirement of supersymmetry)
concerning the geometry of the underlying compactification space 
and the $H$-field configuration,
demand that $X$ is non-K\"ahler 
[\ref{Wit}], cf. also [\ref{tors}], [\ref{torsbecker}], [\ref{CKL}].
The severe consistency problem stems from the fact that
the non-trivial $H$-field (induced from a mismatch between 
$tr R\wedge R$ and $tr F\wedge F$) has actually to be used at the same time
consistently on the right hand side of the anomaly balance for 
the connection $\omega + a H$ from which $tr R\wedge R$ is computed.
Here $\omega$ is the torsion-free spin connection
and  $H$ understood as one-form 
by suitable contractions with vielbeins.
On the level of the three-form $H$ itself this amounts to a cubic relation
$H=dB+\al'\big( \Omega_3(\omega + a H)-\Omega_3(A)\big)$
with the corresponding Chern-Simons terms 
(where $\Omega_3(A)=tr(A\wedge F-\frac{1}{3}A\wedge A\wedge A$)).

Here we are in the case that the Hodge-type of the $H$ field is 
$(2,1)+c.c.$ as $dH$ has to be of type $(2,2)$. (More generally one may
consider also the case of a Hodge type $(3,0)+c.c.$ which could be cancelled
in the complete square part of the Lagrangian by a non-trivial gaugino 
condensate vev [\ref{tors}], [\ref{CKL}]).
This has the consequence that the DUY theorem, assuring for stable bundles
the solvability
of the equations of motion from a topological condition, 
is no longer applicable. Here one has to note that 
for $X$ being non-K\"ahler already the notion of stability
is slightly modified as the would-be K\"ahler form $J$ is no longer closed
with corresponding consequences for $\int_X c_1(V)J^2$ and the notion of slope
[\ref{LY}], [\ref{BM}].
For steps in the direction of a generalisation cf. [\ref{Yau}]
for the case of a $T^2$-fibration with 
base $B$ a $K3$ surface, which from the normal ($H=0$)
perspective would be a rather degenerate exceptional case.\\

\noindent
{\em Corresponding problems in the strongly coupled heterotic setup}

The need to solve the anomaly constraint actually locally 
and not only in the global (cohomological) balance becomes especially 
palpable in the case of heterotic $M$-theory where 
\beqa
(dG)_{\footnotesize{IJKL\, 11}} \sim 
\kappa^{\frac{2}{3}}\Big( \sum_{i=0}^1\, \delta(x_{11}-x_i)
\big( \frac{1}{2}tr\, R_i\wedge R_i - tr \; F_i \wedge F_i\big) 
+16 \pi^2 \sum_{k=1}^m \delta(x_{11}-x_k)\delta_{W_{x_k}}\Big) \wedge dx_{11}
\nonumber
\eeqa
The (spacetime-filling) fivebrane contribution $W$
consists of $m$ components $W_{x_k}$,
supported (compactly) on various (itself not necessarily irreducible)
holomorphic curves $C_k$ 
(of dual four-forms $\delta_{W_k}$ in $X_k$) lying in
the Calabi-Yau space $X_k$ over the point $x_k$ of the interval 
$I=[x_0, x_1]$.
Here the standard embedding
$F_{obs}=R$, ``spin in the gauge'', can no longer be a solution, 
although still fulfilling the global balance.
The reason is that the local character, here along $I$,
of the condition becomes especially pronounced,
demanding that at each end of the interval
one of the individual $E_8$ bundles $V_i$ cancels
the term $\frac{1}{2}tr\; R_i\wedge R_i$.
The idea to represent $W$ as $c_2(V_{hid})$
becomes here obsolete as this strategy is not local in $I$ already, 
let alone locality in $X$.
(For some solutions with $W=0$ locally in $I$, though not in $X$,
cf.~[\ref{LukOvr}].) 

Now, considered locally in $I$, it seems that fivebrane components $W_{x_k}$
can easily be balanced consistently by 
a corresponding $G_{2,2;0}$-contribution\footnote{Boundary 
$G_{2,1;1}$-components of type
$H^{(2,1)} (+c.c.)$ times $\delta(x_{11}-x_0)\wedge dx_{11}$, if possible,
would bring one back to the previous problems so we assume these to be absent
(also with such a contribution
the connection $\omega + a \delta(x_{11}-x_i)H$
used in $1/2 \, tr R_i\wedge R_i$ would have a problematic 
delta-function singularity).
So a potential consistency problem from the connection $\omega + aH$ 
is not induced as $H=0$.
Still the volume modulus can be stabilised
(without $H^{(2,1)}\neq 0$, $X$ non-K\"ahler)
by worldsheet instantons [\ref{CKL}] 
or the S-Track mechanism [\ref{CK S-T}].} 
of the form 
step function $\theta(x_{11}-x_k)$ times the $(2,2)$ form $\delta_{W_{x_k}}$.
Similarly, 
to absorb a mismatch $\frac{1}{2}tr\, R_i\wedge R_i-tr \; F_i \wedge F_i$ 
one can also use a boundary contribution $G=\theta$ times $(2,2)$
which therefore, it seems, considered just locally in $I$,
could absorb both types of contributions from the right hand side.

However, actually the various contributions, locally constant along $x_{11}$,
have to fit together in the ``upstairs picture'' on ${\bf S^1}$
with a ${\bf Z_2}$ action, i.e., 
the various jumps in total have to compensate each other.
As $G$ is odd, jumps in the bulk cancel mutually and the fixpoint
contributions remain.
So, having no $H$-type boundary components in $G$,
one ends up with 
\beqa
\label{G-balance}
0&=& \sum_{i=0}^1\,
\big( \frac{1}{2}tr\, R_i\wedge R_i - tr \; F_i \wedge F_i\big) 
+16 \pi^2 \sum_{x_k\, = \, x_0\; or \; x_1} \delta_{W_{x_k}}
\eeqa
As the smooth and delta-function parts have to cancel
individually, 
one gets (\ref{G-balance}) just for the smooth terms and no fivebranes
on the boundaries (anti-fivebranes being forbidden).

\subsection{Discussion}

Let us emphasize that $W=0$, seemingly avoiding $G\neq 0$, 
is satisfied in the mentioned models [\ref{LukOvr}]
only at the cohomological level. Realizing 
that the condition to be solved is on the 
differential form level one does not get $G=0$ models. 
Similar remarks apply to older $(0,2)$ models
which solve the anomaly without fivebranes but only on the
cohomological level, and to the models
we present in this paper. Nevertheless let us point out two things.
The ability to avoid the delta-function contribution reduces already
substantially the problem of potential inconsistencies when one tries
to solve the anomaly constraint (\ref{form level}) as the problem with
singularities occuring on both sides in different orders is avoided.
The general expectation is then that some relevant features of
the models (understood in the naive $H=0$ sense) persist, 
despite the remaining necessity to
adjust, perhaps order by order, a non-trivial
(but now smooth) $H$-field configuration. But note 
the caveat that the radius will be fixed to a finite value, so there 
will be no large radius limit for a perturbative supergravity treatment;
also a stability notion for $X$ non-K\"ahler is more subtle [\ref{BM}]. 
The same philosophy underlies procedures to check that 
the number of fivebranes wrapped on elliptic fibers
in heterotic spectral cover models matches the number of
threebranes in a dual $F$-theory model [\ref{FMW}], [\ref{AC5=3}];
or also the numerous phenomenological studies of heterotic compactifications
done so far.
The other point, specific to our $W=0$ models, is that they live still on
elliptically fibered Calabi-Yau spaces. For many investigations touching
more conceptual questions (dualities with other string models is a 
prominent example) these seemingly more abstract 
compactifications have turned out to be more
suitable than Calabi-Yau spaces given by embeddings
in a weighted projective space (or products of them).\\

In {\em section $2$} we discuss 
the various options for the (bases $B$ of the elliptically fibered)
spaces over which we build our bundles.
In {\em section $3$} we recall some notions related to stability and describe 
in {\em section $4$} the way we build our bundles
which enable us to get $W=0$ (this contrasts with the fact [\ref{FMW}]
that bundles built with the spectral cover method give $W\neq 0$). There
the stability of the bundles could be established more directly
[\ref{FMWIII}] whereas here the issue of stability becomes a major
technical point to which we do full justice only in the mathematical
companion paper [\ref{math paper}]. 
In {\em section $5$} we give for $B$ an Enriques surface
the relevant stability results
and show that within our class of bundles $W\geq 0$ is violated.
Stability results and $W=0$ examples for other bases are given 
in {\em section $6$}.

\section{The elliptic Calabi-Yau space over various bases}

\resetcounter

As our procedure to build bundles on $X$ involves an extension
of bundles $U_p=\pi^* E_p$ which are themselves pull-backs of bundles $E_p$ 
on the base $B$
one has to ensure the stability of such pullback bundles $U_p$ if 
$E_p$ is stable. For this we distuinguish two cases among our set
of bases, consisting of Hirzebruch surfaces ${\bf F_r}$ ($r=0,1,2$), 
del Pezzo surfaces $dP_k$ ($k=0, \dots, 8$)
and the Enriques surface $E$.
The search for a specific gauge group,
be it of a GUT theory or of the Standard model,
motivates the choice of a Hirzebruch surface or the Enriques surface as base.
On the one hand we will treat the Enriques surface $E$
which is also of special physical importance as a GUT group $SU(5)$ 
there can be broken to the Standard Model group because of 
$\pi_1(X_E)={\bf Z_2}$. On the other hand we will treat in section
\ref{ample antiK}
the case where the anticanonical bundles is ample, i.e. the remaining cases
except ${\bf F_2}$.

\subsection{The case of $B$ an Enriques surface}

Let us describe the different physical and mathematical issues related to the 
use of the Enriques surface as base.

\subsubsection{On the physical motivation for the Enriques surface}

One approach to realize the Standard model gauge 
group within heterotic string compactifications is to build a bundle $V$
of structure group $SU(5)$, leading to an unbroken gauge group given by
the grand unified group $SU(5)$ in the observable sector.
In that case $\pi_1(X)={\bf Z_2}$ allows for a 
Wilson line, breaking the commutator $SU(5)$ (in $E_8$) of a
structure group $SU(5)$ to the Standard model gauge group.
For this one needs
a Calabi-Yau space whose non-trivial fundamental group contains a ${\bf Z}_2$.
$X$ is non-simply connected only if the base $B$ of the fibration
is given by an Enriques surface $E$ where $\pi_1(X)=\pi_1(B)={\bf Z}_2$
(where $c_1:=c_1(B)$ is a two-torsion class).

This approach is in contrast to the procedure where one starts
from a simply connected
Calabi-Yau threefold $X'$ having a free involution $\tau$
from which the required non-simply connected Calabi-Yau space $X$
is built as $X=X'/{\bf Z}_2$. The existence of $\tau$ is related
to the existence of a second section of the elliptic fibration
[\ref{DOPWI}], [\ref{DOPWII}], [\ref{DOPWIII}], [\ref{BD}], [\ref{BCD}],
[\ref{B fibre I}], [\ref{B fibre II}], [\ref{B fibre III}].
Then one searches for $\tau$-invariant bundles having $6$ generations.

By contrast, in the case of the Enriques Calabi-Yau,
one searches directly on $X$
for bundles of net generation number $N_{gen}=\pm 3$.
This, however, led for spectral bundles to the following problem. 
From the mismatch of anomaly cancellation between
$c_2(V)$ and $c_2(X)$ one has to introduce a number of five-branes of
total cohomology class $W=w_B\si+a_f F$.
One has the effectivity condition $w_B=12c_1-\eta=-\eta\geq 0$ where
$\eta\in H^2(B, {\bf Z})$ is a datum describing the bundle 
(the spectral surface
$C$ has cohomology class $n\si+\eta$ where $n$ is the rank of $V$).
$\eta$ must be an effective class satisfying $\eta\geq nc_1$
which for $n$ even reduces to $\eta\geq 0$ and for $n$ odd
to $\eta\geq c_1$; so $\eta=0$ or $c_1$,
giving $N_{gen}=\la\eta(\eta-nc_1)=0$.
(For another attempt cf. [\ref{Thomas}].)

\subsubsection{Mathematical details on the Enriques surface}

\noindent
Consider standard (fibre type $A$)
elliptically fibered CY spaces $X$ with one section
[\ref{FMW}]
and base given by an Enriques surface [\ref{CoDo}], i.e.,
$h^{1,0}(B)=0$ and $K_B^2={\cal O}_B$. $B$ has non-trivial Hodge numbers
$h^{1,1}=10, h^{0,0}=h^{2,2}=1$ , so $c_1^2=0$
and $c_2=12$, 
and middle cohomology
\beqa
H^2(B,{\bf Z})={\bf Z}^{10}\oplus{\bf Z}_2\;\;\;\;\; \mbox{with} \;
\mbox{intersection} \;\mbox{lattice}\;\;\;
\Gamma^{1,1}\oplus \, E_8^{(-)}=\footnotesize{\left( \begin{array}{cc}
0 & 1 \\ 1 & 0 \end{array} \right)}\oplus \, E_8^{(-)}
\eeqa
(orthogonal decompositions). 
Further $\phi c_1=0$ for all $\phi \in H^2(B, {\bf Z})$.
$B$ is always elliptically fibered with fibre $f$ over $b={\bf P^1}$.
However two of the fibers, $f_1$ and $f_2$, are double fibers:
$f=2f_i$, which prevents $B$ from having a section and
$c_1 = f_1 - f_2$ is not effective.

On a generic ('unnodal') $B$ no smooth rational curves exist and
all irreducible curves $C$ have $C^2\geq 0$.
The integral classes in one of the two components
of the cone in $H^2(B, {\bf R})$
defined by $C^2\geq 0$ constitute the effective cone (adding
the torsion class $c_1$ does not matter for this if $C\neq 0$).
For $C$ nef (i.e., $DC\geq 0$ for all curves $C$ on $B$)
$|C|$ is base-point-free, $C$ is ample
if also $C^2\geq 6$ [\ref{CoDo}]. $C=xa+yf=(x,y) \in \Gamma^{1,1}$ is nef 
for $x,y\geq 0$.\\

$B$ can be represented as the qoutient of a $K3$ surface
by a free involution. 
The corresponding $\pi_1(B)={\bf Z}_2$ is inherited by
the elliptic Calabi-Yau space $X$ which itself is
a quotient by a free involution on $K3\x T^2$ (also acting
as $z\ra -z$ on the $T^2$).
The holomorphic two-form $\Omega_2$ of $K3$ being odd, the holomorphic
three-form $\Omega_2\wedge dz$ is preserved, the quotient $X$
being a Calabi-Yau space of vanishing Euler number.

\section{The condition of stability}

\resetcounter

We will choose as polarization $J=z\si + \pi^* H$
where $H$ (chosen in the integral cohomology)
is in the K\"ahler cone $\C_B$ of the base $B$ and $z\in {\bf R^{>0}}$.
For an elliptically fibered Calabi-Yau space $X$ one has
that $J$ is a K\"ahler class if [\ref{B fibre III}]
\beqa
\label{C_X criterion}
J\in \C_X & \Longleftrightarrow & z>0 \; , \;\; H - z c_1 \in \C_B
\eeqa
Stability of a bundle $V$ (with respect to $J$) means $\mu_J(V')< \mu(V)$
for all coherent subsheafes $V'$ of $V$ of $rk \, V' \neq 0, rk \, V$.
Here $\mu(V)= \frac{1}{rk\, V} \int c_1(V) J^2$ is the slope of $V$ 
with respect to $J$.
Similarly semistability is defined by the condition $\mu_J(V')\leq 
\mu(V)$.

A bundle $V$ stable w.r.t.~$J$ satisifes the Bogomolov inequality
\beqa
c_2(V)J \geq 0
\eeqa
The decomposition of the fivebrane class
$W=w_B \si + a_f F = c_2(X)-c_2(V)$ becomes for $B$ an Enriques surface
$12F-c_2(V)$ which gives then
\beqa
c_2(V)J = -w_B H + z(12-a_f)
\eeqa
For bundles built by the spectral cover construction
one knows by [\ref{FMWIII}], Thm.~7.1 
that suitable $J$ must have $z$ sufficiently small 
({\em ``spectral polarizations''}).
But if $z$ has to be chosen negligible small,
only $w_B=0$ is possible as $-w_B H\leq 0$ by the requirement $w_B\geq 0$.
The latter stems from the condition that the fivebrane class is effective.
On the other hand, below we will assure stability
for $z$ sufficiently {\em large}; note that on $B$ the Enriques surface
there will be no problem then with the criterion (\ref{C_X criterion})
when just chosing $H\in \C_B$.

Let us consider now the stability of zero-slope bundles $V$ constructed as 
extensions 
\beqa
\label{nonsplit}
0 \to U \to V \to W \to 0
\eeqa
Here $U$ and $W$ are assumed to be stable.
Necessary conditions for the stability of $V$ are that
\begin{itemize}
\item $\mu(U)<0$ 
\item the $W$ of $\mu(W)>0$ is not a 
subbundle of $V$, i.e., the extension (\ref{nonsplit}) is non-split
\end{itemize}

\section{Bundles built as extensions}

\resetcounter

As mentioned earlier one problem in heterotic model building,
especially on elliptically fibered Calabi-Yau spaces $X$,
is the occurrence of a number of space-time filling fivebranes [\ref{FMW}], 
preventing the model to be interpreted as a perturbative nonlinear sigma model.
More specifically, this problem occurs within the spectral cover construction
[\ref{FMW}, \ref{don1}] (equivalently understood as a relative Fourier-Mukai 
transform [\ref{AH1}]). The advantage of this method is an improved flexibility
in the explicitely computed net generation number [\ref{C}], [\ref{bjoern}].

This problem is circumvented in the present paper as follows. We define a 
stable $SU(n)$ bundle $V$ as a non-trivial extension of bundles $U$ and $W$ 
of lower rank, especially when $W$ is a line bundle. The question of stability,
and already the existence of a non-split extension, turns out to be 
non-trivial. The bundles $U$ and $W$ are constructed as pull-backs from the 
base $B$, twisted by certain line bundles. For $B$ a Hirzebruch surface 
${\bf F_r}$ we find GUT models with chiral matter, and for $B$ the Enriques 
surface the Standard Model gauge group. In the GUT case it is possible to 
avoid fivebranes in the anomaly constraint. Thereby one stays within the 
framework of perturbative $(0,2)$ models, characterised by stable holomorphic 
bundles $V$ of $c_1(V)=0$ (this condition can be relaxed) satisfying
\beqa
c_2(V)=c_2(TX)
\eeqa

Now, in the spectral cover construction the bundle decomposes on the 
generic fibre $F$ as 
\beqa
V|_F=\bigoplus_{i=0}^n \cO_F(p_i-p_0)\;\;\;\;\;\; \mbox{with} \;\; \sum p_i=p_0
\eeqa
and is adiabatically extended along the base: the system of the $(p_i)$ becomes
base-point dependent, leading to a $n$-fold cover $C$ of $B$, including
a twist by a line bundle $\cO_B(\eta)$ on $B$.

A different starting point is to choose $V$ on $F$ as
\beqa
\label{on the fibre}
V|_F=\bigoplus \cO_F(x_i p_0)\;\;\;\;\;\; \mbox{with} \;\; \sum x_i=0
\eeqa
One of the simplest possibilities is to choose $x_i=:-x$ for $i=1, 
\dots , n$ and $x_0=nx$. 
Thus, rewriting (\ref{on the fibre}),
one starts from a {\em split} short exact sequence 
\beqa
0\lra \cO_F^n\ox \cO_F(-xp_0)\lra V|_F \lra \cO_F(nxp_0)\lra 0
\eeqa
To spread this out along $B$
one chooses now $V_{n+1}$ as the extension
\beqa
0 \lra \pi^* E_n \ox \cO_X\big( - x\si -\pi^*\al\big) \lra V_{n+1}\lra 
\cO_X\big(n(x\si + \pi^*\al)\big) \lra 0
\eeqa
Here the non-trivial information about the bundle along 
the direction $B$ is encoded in $E_n$, including
a twist by a line bundle $\cO_B(\al)$ of the base.
More generally, one may consider the case where the set of the $x_i$ 
partitions into two sets of $x_i=px$ for $q$ $i$'s and $x_i=-qx$ for $p$
$i$'s, which globally corresponds to the extension
(with $U_p=\pi^* E_p, W_q=\pi^* E_q$ and $D=x\si +\pi^*\al$)
\beqa
\label{defining sequence}
0 \lra  U_p\ox \cO_X(-qD)\lra  V_{p+q} 
\lra W_q \ox\cO_X(pD) \lra  0
\eeqa
(where one demands $DJ^2>0$ so that the slope condition
$\mu(U_p\ox \cO(-qD))<0$ is fulfilled).
So the class of bundles we consider 
is given by bundles $V_{p+q}$ of rank $p+q$ defined as non-trivial 
extensions (\ref{defining sequence})
of stable bundles $U_p$ and $W_q$ of rank $p$ and $q$
with $c_1(U_p)=0=c_1(W_q)$, 
suitably twisted by powers of a line bundle $\cO(D)$
so as to preserve $c_1(V_{p+q})=0$.

One can consider in particular the case that $U_p$ and $W_q$
are pull-backs $\pi^* E_p$, $\pi^* E_q$ of bundles on $B$. Then 
$\pi^*E$ is (semi-)stable if $E$ is (semi-)stable on $B$ 
with respect to $H$, say for $B$ the Enriques surface;
the detailed arguments for this 
and all mathematical statements concerning the stability proofs and 
non-split conditions below are given in the 
mathematical companion paper [\ref{math paper}]. Later we will 
actually show stability of $V_{p+q}$ only for $q=1$.

One finds with $c_2(U_p)=uF$ and $c_2(W_q)=wF$ 
\beqa
c_2(V_{p+q})&=&-\frac{1}{2}pq(p+q)D^2+(u+w)F
\eeqa

Let us now come to the physical conditions.
One must ensure the effectivity of the class
\beqa 
W=w_B\si +a_f F=c_2(X)-c_2(V_1)-c_2(V_2)
\eeqa 
of the fivebrane.
Assuming for simplicity no hidden sector bundle
one finds as components 
\beqa 
w_B=12 c_1 + \frac{1}{2}pq(p+q)x(2\al-xc_1)& , &
a_f=c_2+11c_1^2 + \frac{1}{2}pq(p+q)\al^2 - (u+w)\;\;\;
\eeqa

For the net chiral matter content one finds as generation number 
\beqa
N_{gen}=\frac{1}{2}c_3(V_{p+q})&=&\frac{pq}{6}(p^2-q^2)D^3+
x( qu-pw)+\frac{1}{2}c_3(U_p)+\frac{1}{2}c_3(W_q)
\eeqa
Note that in the special case that $U_p$ and $W_q$ are pullbacks
from the base (so that they have vanishing third Chern class)
one finds $N_{gen}\sim x$ over the Enriques base.

Below we will prove stability of $V_{p+q}$ for $q=1$
given a stable bundle $U_p$. To get a concrete stable bundle $U_p$
we take $U_p=\pi^* E_p$. 
Over $B$ with $K_B^{-1}$ ample one gets then GUT models
with $W=0$ and $N_{gen}\neq 0$ (cf. section \ref{ample antiK}).
Over the Enriques base one can get the 
Standard model gauge group; then, however, one encounters 
the side effect that $N_{gen}$ 
will run just with $x$ as $w_B$ does; furthermore, only 
the case $w_B=0$ would be allowed (as will be seen below), giving $x=0$; 
for $x=0$ stability, however, cannot be assured.

\section{Standard model groups: Enriques base}

\resetcounter

For the case $(p,q)=(n,1)$ $V_{n+1}$ can be shown to be stable. So
let $V_n$ be a stable  bundle of $c_1(V_n)=0$ and
$D=x\si + \pi^* \al$ and define $V_{n+1}$ as a non-split extension
(here $W_1=\cO$)
\beqa
\label{n+1 extension}
0\to V_n\ox \cO(-D)\to V_{n+1}\to \cO(nD)\to 0
\eeqa
Let us first discuss the non-split condition.
For this assume $V_n=\pi^*E_n$ with $c_1(E_n)=0$ and $E_n$ stable. Then
one finds for $x>0$ 
and for $a:=\al H<0$ the index condition [\ref{math paper}]
\beqa
I:=n-c_2(E)+\frac{n(n+1)^2}{2}\al^2>0
\eeqa
for precisely when a non-split extension exists.
If $x\leq 0$ then $Ext^1\neq 0$ exactly if $I<0$.\\

\noindent
{\em Stability of $V_{n+1}$ and physical constraints}\\

We note the following necessary condition: 
if $V_{n+1}$ is stable (so (\ref{n+1 extension}) is non-split) then
\beqa
\label{necessary condition}
x\neq 0&\Longrightarrow & x \cdot a <0
\eeqa
where $a:= \al H$.
$V_{n+1}$ has now specific stability regimes  
w.r.t.~the K\"ahler class $J=z\sigma+\pi^*H$
\beqa
0\; <x< -a &\lra &\;\;\; 
\frac{n x}{-na+1}\, \frac{H^2}{2}
\;\;< \;\; z \;\; < \;\; \frac{nx}{-na}\, \frac{H^2}{2}\\
-a\; <x\; <\;\; 0 & \lra & \;\;\;\;\;\;\;\;\;\;\;\frac{nx}{-na}\, 
\frac{H^2}{2}\;\; <\;\; z
\;\; <\;\; \frac{nx}{-na+1}\, \frac{H^2}{2}
\eeqa

The physical constraints concern the effectivity of the fivebrane
$W=w_B\si + a_f F$ where
\beqa
w_B=\frac{1}{2}n(n+1)x(2\al-xc_1)\geq 0 & , & 
a_f=12+\frac{1}{2}n(n+1)\al^2- c_2(E) \geq 0
\eeqa
and the phenomenological value  $\pm 3$ of the net generation number
\beqa
N_{gen}=x\Big( \frac{1}{2}n(n^2-1)\al^2 + c_2(E)\Big)
\eeqa
Note that, as $w_B\geq 0$ requires therefore $x\al\geq 0$, one gets 
in view of (\ref{necessary condition}) 
\beqa
w_B\geq 0\Longrightarrow x=0
\eeqa
If a hidden sector bundle of the same type
is turned on the argument remains 
valid as $w_B=\sum_{i=1}^2 c_i\, x_i\al_i\geq 0$
(with $c_i>0, c_1=\frac{n(n+1)}{2}$) gives $w_B H\geq 0$, 
a contradiction to (\ref{necessary condition}).

$x=0$ is the case for which the existence of stable bundles could not be
assured above.

\section{\label{ample antiK}GUT groups: working over $B$ with ample $K_B^{-1}$}

\resetcounter

In  this section $B$ denotes a surface with ample $K_B^{-1}$.
Now $\pi^*E$ is (semi-)stable on $X$ with respect to $J=z\si +\pi^*H\in \C_X$ 
if $E$ is (semi-)stable on $B$ with respect to $H=hc_1$;
here $H-zc_1\in \C_B$ gives $z<h$.
Given the fact that $c_1$ is now no longer a two-torsion class,
a greater numerical freedom between $W$ and $N_{gen}$ occurs, in
particular the common proportionality to the parameter $x$ is lifted.
Therefore it is possible here to have $W=0$ and $N_{gen}\neq 0$.

Concretely the fivebrane class has components
(assuming for simplicity $V_{hid}=0$)
\beqa 
w_B = 12c_1 + \frac{1}{2}n(n+1)x(2\al-xc_1)\;\;\;\;\; , \;\;\;\;\;
a_f = c_2   + \frac{1}{2}n(n+1)\al^2 - c_2(E) + 11c_1^2
\eeqa

In contrast to the case of the Enriques base it is now possible 
to satisfy $w_B\geq 0$ while having $x\neq 0$.
One finds now $W=0$
(just to get $w_B, a_f\geq 0$ is easy)
for the choices
\beqa
\al=\Big(\frac{x^2}{2}-\frac{12}{n(n+1)}\Big)\frac{c_1}{x}
& \Longrightarrow  & w_B=0\\
c_2(E)=c_2+11c_1^2+\frac{n(n+1)}{2}\al^2
& \Longrightarrow  & a_f=0
\eeqa

For instance, for building an $SO(10)$ GUT model without fivebranes one can use
the twist $D=\si -\pi^*c_1/2$ and a rank $n=3$ bundle $E$
on a base ${\bf F_r}$ of instanton number $104$. Or one may 
construct an $E_6$ GUT model without fivebranes
from using the twist $D=2\si$ and a plane bundle of $c_2(E)=92$.
(One immediately checks the non-split conditions.)

One gets furthermore that
\beqa
N_{gen}=x\Big( \frac{n(n^2-1)}{6}\big(3\al^2-3x\al c_1+x^2c_1^2\big) 
+ c_2(E)\Big)
\eeqa
\\

Let us mention that one can carry through a similar program
also for extensions by spectral bundles (without special restrictions 
on the base surface $B$), leading to examples
of stable bundles without five-branes, cf.~[\ref{math paper}].

So the general lesson in all the different cases is similar: the greater
numerical freedom provided by the twist and the extension can allow one to have
$W=0$, the burden then is however to prove stability of the extensions
[\ref{math paper}].\\

We thank H. Kurke, A. Krause and D. Hern\'andez Ruip\'erez for discussion.

\section*{References}
\begin{enumerate}

\item
\label{Wit}
E. Witten, 
{\em New Issues in manifolds of SU(3) Holonomy}, 
Nucl. Phys {\bf B268} (1986) 79.\\
M. Green, J. Schwarz and E. Witten,
{\em Introduction to Superstring Theory, Vol. II},
Cambridge University Press (1988).\\
A. Strominger, {\em Superstrings with Torsion}, 
Nucl. Phys. {\bf B274} (1986) 253.

\item
\label{LY}
J. Li and S.T. Yau,
{\em  Hermitian Yang-Mills connections on non-K\"ahler manifolds}, 
in {\em ``Math.
aspects of string theory''} (S.-T. Yau ed.), World Scientific Publ. 1987.

\item
\label{FMW}
R. Friedman, J.W. Morgan and E. Witten,
{\em Vector Bundles And F Theory},
hep-th/9701162, Commun.Math.Phys. {\bf 187} (1997) 679.

\item
\label{don1}
R. Donagi, {\em Principal bundles on elliptic fibrations}, alg-geom/9702002,
Asian J. Math. {\bf 1} (1997), 214.

\item
\label{C}
G. Curio, {\em Chiral matter and transitions in heterotic string models},
hep-th/9803224, Phys.Lett. {\bf B435} (1998) 39.

\item
\label{bjoern}
B. Andreas,
{\em On Vector Bundles and Chiral Matter in N=1 Heterotic Compactifications},
hep-th/9802202, JHEP {\bf 9901} (1999) 011.

\item
\label{AH1}
 B. Andreas and  D. Hern\'andez Ruip\'erez, 
{\em U(n) Vector Bundles on Calabi-Yau Threefolds
for String Theory Compactifications}, hep-th/0410170, 
Adv. Theor. Math. Phys. 9: 253, 2006.\\
B. Andreas and  D. Hern\'andez Ruip\'erez, 
{\em Fourier-Mukai Transforms and Applications to String Theory}, 
math.AG/0412328, 
Rev. R. Acad. Cien. Serie A. Mat. Vol. 99 (1), 2005, 29-77.\\
B. Andreas,
{\em The Fourier-Mukai transform in string theory}, hep-th/0505263,
Encyclopedia of Mathematical Physics, eds. J.-P. Francoise, 
G. L. Naber and Tsou S. T., Oxford: Elsevier, 2006.

\item
\label{db}
B. Andreas and D.Hernandez Ruiperez,
{\em Comments on N=1 Heterotic String Vacua},
hep-th/0305123, Adv.Theor.Math.Phys. {\bf 7} (2004) 751.

\item
\label{ov}
B.A. Ovrut, T. Pantev and J. Park,
{\em Small Instanton Transitions in Heterotic M-Theory},
hep-th/0001133, JHEP 0005 (2000) 045.

\item
\label{FMWIII}
R. Friedman, J.W. Morgan and E. Witten,
{\em Vector Bundles over Elliptic Fibrations}, alg-geom/9709029,
Jour. Alg. Geom. {\bf 8} (1999) 279.

\item
\label{DOPWI}
R. Donagi, B. Ovrut, T. Pantev and D. Waldram,
{\em Standard-Model Bundles on Non-Simply Connected Calabi--Yau Threefolds},
hep-th/0008008, JHEP {\bf 0108} (2001) 053.

\item
\label{DOPWII}
R. Donagi, B. Ovrut, T. Pantev and D. Waldram,
{\em Standard-model bundles},
math.AG/0008010, Adv.Theor.Math.Phys. {\bf 5} (2002) 563.

\item
\label{DOPWIII}
R. Donagi, B. Ovrut, T. Pantev and D. Waldram,
{\em Spectral involutions on rational elliptic surfaces},
math.AG/0008011, Adv.Theor.Math.Phys. {\bf 5} (2002) 499.

\item
\label{BD}
V. Bouchard and R. Donagi,
{\em An SU(5) Heterotic Standard Model},
Phys.Lett. {\bf B633} (2006) 783,
hep-th/0512149.

\item
\label{BCD}
V. Bouchard, M Cvetic and R. Donagi,
{\em Tri-linear Couplings in an
Heterotic Minimal Supersymmetric Standard Model},
hep-th/0602096.

\item
\label{B fibre I}
B. Andreas, G. Curio and A. Klemm,
{\em Towards the Standard Model spectrum from elliptic Calabi-Yau},
hep-th/9903052, Int.J.Mod.Phys. {\bf A19} (2004) 1987.

\item
\label{B fibre II}
G. Curio,
{\em Standard Model bundles of the heterotic string},
hep-th/0412182, Int.J.Mod.Phys. {\bf A21} (2006) 1261.

\item
\label{B fibre III}
B. Andreas and G. Curio,
{\em Standard Models from Heterotic String Theory},
hep-th/0602247,



\item
\label{Thomas}
R.P. Thomas, {\em An obstructed bundle on a Calabi-Yau 3-fold}, Adv. 
Theor. Math. Phys {\bf 3} (1999) 567,math.AG/9904034.\\
R.P. Thomas, {\em Examples of bundles on Calabi-Yau 3-folds for string theory 
compactifications},
Adv.Theor.Math.Phys. {\bf 4} (2000) 231, math.AG/9912179.

\item
\label{CoDo}
F. Cossec and I. Dolgachev, {\em Enriques Surfaces I}, Birkh\"auser 1989.

\item
\label{tors}
G. L. Cardoso, G. Curio, G. Dall'Agata and D. Lust,
{\em Heterotic String Theory on non-Kaehler Manifolds with H-Flux 
and Gaugino Condensate},
hep-th/0310021, Fortsch.Phys. {\bf 52} (2004) 483.\\
G. L. Cardoso, G. Curio, G. Dall'Agata and D. Lust,
{\em BPS Action and Superpotential for Heterotic String Compactifications 
with Fluxes}, hep-th/0306088, JHEP {\bf 0310} (2003) 004.\\
G. L. Cardoso, G. Curio, G. Dall'Agata, D. Lust, P. Manousselis and 
G. Zoupanos,
{\em Non-Kaehler String Backgrounds and their Five Torsion Classes},
hep-th/0211118, Nucl.Phys. {\bf B652} (2003) 5.

\item
\label{torsbecker}
K. Becker, M. Becker, K. Dasgupta and S. Prokushkin,
{\em Properties Of Heterotic Vacua From Superpotentials},
hep-th/0304001, Nucl.Phys. {\bf B666} (2003) 144.\\
K. Becker, M. Becker, K. Dasgupta and P.S. Green,
{\em Compactifications of Heterotic Theory on Non-Kahler Complex Manifolds: I},
hep-th/0301161, JHEP {\bf 0304} (2003) 007.\\
K. Becker, M. Becker, K. Dasgupta, P.S. Green and E. Sharpe,
{\em Compactifications of Heterotic Strings on Non-Kahler 
Complex Manifolds: II}, 
hep-th/0310058, Nucl.Phys. {\bf B678} (2004) 19.

\item
\label{CKL}
G. Curio, A. Krause and D. Lust,
{\em Moduli Stabilization in the Heterotic/IIB Discretuum},
hep-th/0502168, Fortsch.Phys. {\bf 54} (2006) 225.

\item
\label{CK S-T}
G. Curio and A. Krause,
{\em S-Track Stabilization of Heterotic de Sitter Vacua},
hep-th/0606243.

\item
\label{Douglas}
M.R. Douglas, R. Reinbacher and S.-T. Yau,
{\em Branes, Bundles and Attractors: Bogomolov and Beyond},
math.AG/0604597.

\item
\label{Yau}
J.-X. Fu and S.-T. Yau,
{\em The theory of superstring with flux on non-Kahler manifolds 
and the complex Monge-Ampere equation},
hep-th/0604063.\\
K. Becker, M. Becker, J.-X. Fu, L.-S. Tseng and S.-T. Yau,
{\em Anomaly Cancellation and Smooth Non-Kahler Solutions 
in Heterotic String Theory},
hep-th/0604137, Nucl.Phys. {\bf B751} (2006) 108.

\item
\label{LukOvr}
A. Lukas and B.A. Ovrut,
{\em Symmetric Vacua in Heterotic M-Theory},
hep-th/9908100.

\item
\label{math paper}
B. Andreas and G. Curio,
{\em Stable bundle extensions on elliptic Calabi-Yau threefolds},
math.AG/0611762.

\item
\label{BM}
V. Brinzanescu and R. Moraru,
{\em Holomorphic rank-2 vector bundles on non-Kahler elliptic surfaces},
math.AG/0306191;\\
V. Brinzanescu and R. Moraru,
{\em Stable bundles on non-Kahler elliptic surfaces}, math.AG/0306192.

\item
\label{DZ}
M.R. Douglas and C.G. Zhou,
{\em Chirality Change in String Theory},
hep-th/0403018, JHEP {\bf 0406} (2004) 014.\\
M.C. Brambilla, 
{\em Semistability of certain bundles on a quintic Calabi-Yau threefold},
math.AG/0509599.

\item
\label{AC5=3}
B. Andreas and G. Curio,
{\em Three-Branes and Five-Branes in N=1 Dual String Pairs},
hep-th/9706093, Phys.Lett. {\bf B417} (1998) 41.\\
B. Andreas and G. Curio,
{\em On discrete Twist and Four-Flux in N=1 heterotic/F-theory 
compactifications},
hep-th/9908193, Adv.Theor.Math.Phys. {\bf 3} (1999) 1325.\\
B. Andreas and G. Curio,
{\em Horizontal and Vertical Five-Branes in Heterotic/F-Theory Duality},
hep-th/9912025, JHEP {\bf 0001} (2000) 013.

\item
\label{Kim}
H. Kim, 
{\em Moduli Spaces of Stable Vector Bundles on Enriques Surfaces},
Nagoya Math. J., Vol. {\bf 150} (1998) 85.

\end{enumerate}

\end{document}